# Onboard Calibration Circuit for the Front-end Electronics of DAMPE BGO Calorimeter*

ZHANG De-Liang[1,2]   FENG Chang-Qing[1,2;1)]   ZHANG Jun-Bin[1,2]   WANG Qi[1,2]
MA Si-Yuan[1,2]   GAO Shan-Shan[1,2]   SHEN Zhong-Tao[1,2]   JIANG Di[1,2]
GUO Jian-Hua[3]   LIU Shu-Bin[1,2]   AN Qi[1,2]

[1]Department of Modern Physics, University of Science and Technology of China, Hefei 230026, China
[2]State Key Laboratory of Particle Detection and Electronics (IHEP-USTC), Hefei 230026, China
[3]Purple Mountain Observatory, Chinese Academy of Sciences, Nanjing 210008, China

**Abstract:** The DAMPE (DArk Matter Particle Explorer) is a scientific satellite mainly aiming at indirectly searching for dark matter in space. One critical sub-detector of the DAMPE payload is the BGO (Bismuth Germanate Oxid) Calorimeter, which contains 1848 PMT dynode signals and 16 FEE (Front-End Electronics) boards. VA160 and VATA160, two 32-channel low power ASIC (Application Specific Integrated Circuit) chips, are adopted as the key components on the FEEs to perform charge measurement for the PMT signals. In order to monitor the parameter drift which may be caused by temperature variation, aging, or other environmental factors, an onboard calibration circuit is designed for the VA160 and VATA160 ASICs. It is mainly composed of a 12-bit DAC (Digital to Analog Converter), an operation amplifier and an analog switch. Test results showed that a dynamic range of 0~30 pC with a precision of 5 fC (RMS) was achieved, which covers the VA160's input range. Furthermore, it is used to test the trigger function of the FEEs. The calibration circuit has been implemented for the front-end electronics of BGO Calorimeter and verified by all the environmental tests for both Qualification Model and Flight Model of DAMPE. The DAMPE satellite will be launched at the end of 2015 and the calibration circuit will perform onboard calibration in space.
**Keywords:** DAMPE, BGO Calorimeter, Front-end electronics, Calibration circuit, VA160
**PACS:** 29.85.Ca

## 1 Introduction

The DAMPE is a scientific satellite being developed in China, which is planned to operate in a near-earth orbit with an altitude of 500 km for a mission period of at least 3 years. Its main scientific objective is searching for the dark matter particles through obtaining the e+/e- and γ energy spectrum in space, which is regarded to be closely related to the clue of darkmatter [1] [2].

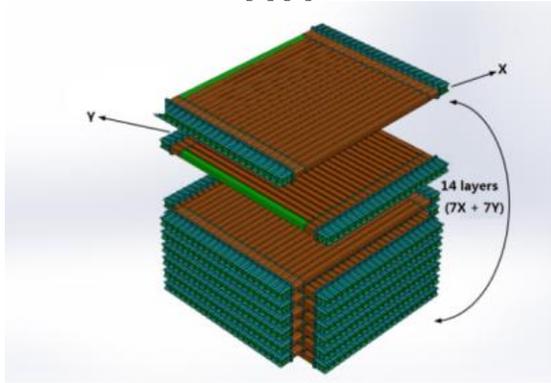

Fig. 1. Crystal bars of BGO Calorimeter

As a critical sub-detector of the DAMPE payload, the BGO Calorimeter is responsible for precisely measuring the energy of cosmic rays from 5 GeV to 10 TeV, distinguishing positrons/electrons and gamma rays from hadron backgrounds, and providing trigger information for the whole DAMPE payload [3] [4].

The BGO Calorimeter contains 308 BGO crystal bars and 616 PMTs. In order to achieve a large dynamic range to $2\times10^5$ for each BGO bar, each of the 616 PMTs in the BGO Calorimeter incorporates a three dynode (2, 5, 8) pick off, which results in 1848 signal channels in total. The read-out electronics contains 16 front-end electronics boards (FEE) and 1848 electronics channels where each electronics channel deals with one dynode signal, measuring charge of the dynode signal. Each electronics channel is required to cover a dynamic range of 0~12pC with a precision better than 10fC and a nonlinearity less than 2% [5] [6] [7].

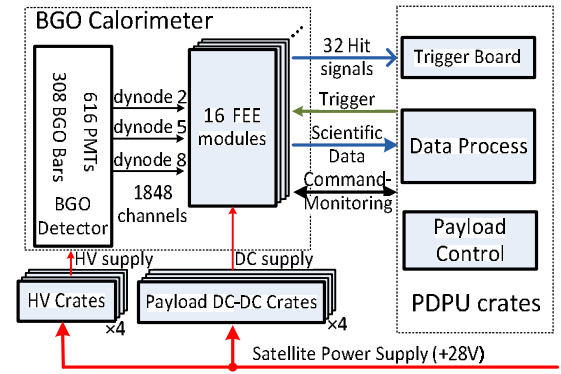

Fig. 2. The readout electronics for BGO Calorimeter

*Supported by the Strategic Priority Research Program on Space Science of Chinese Academy of Sciences (Grant No. XDA04040202-4), and the National Basic Research Program (973 Program) of China (Grant No. 2010CB833002).
1) E-mail: fengcq@ustc.edu.cn

Before launching, the BGO Calorimeter Flight Model should take nearly one year in total to go through a series of functional tests and environmental tests, including EMC (Electromagnetic Compatibility) test, vibration test, thermal cycling test, thermal-vacuum test, 360-hour burn-in test, and later a series of integration level tests together with the satellite platform. During the ground-based tests, we need to monitor the performance of every readout channel of the front-end electronics. More importantly, the DAMPE satellite will continuously operate in orbit for more than 3 years in space, during which the parameters of its front-end electronics may vary or degrade due to aging, temperature drifting, total radiation dose, etc [8].

Therefore onboard calibration for the front-end electronics should be carried out periodically to monitor the performance, diagnose the parameters of the readout channels, or even futher, to compensate the parameter distortion.

## 2 Architecture of the Front-End Electronics

Three types of FEE boards (FEE-A, FEE-B, FEE-C) are designed with the same schematic except for different numbers of channels and FEE-A's additional function of generating hit signals. The FEE boards are configured with the BGO crystal layers as shown in Fig. 3. FEE-A and FEE-B contain 144 electronics channels which are responsible for acquiring 132 dynode signals from two crystal layers and FEE-C contains 72 electronics channels which are responsible for acquiring 66 dynode signals from only one crystal layer. The other channels, which are not connected to PMTs, are just spare ones used as references.

FEE-A's schematic is shown in Fig. 4. It contains several parts, charge measurement circuit, calibration circuit, hit signal generating circuit (only in FEE-A), current and temperature monitoring circuit, power supply circuit and communication circuit between FEE and PDPU (Payload Data Process Unit), which are all controlled by an Actel Flash FPGA [9]. Among them, the charge measurement circuit is the critical part, which is responsible for acquiring the detector's dynode signals. It's made up of 2 VA160s, 4 VATA160s, two operational amplifiers and a high-precision ADC (Analog to Digital Converter). By receiving current pulse signals from PMT dynodes, integrating the charge in VA160 or VATA160, amplifying through 2 level operational amplifiers and digitizing with the ADC, FEE obtains the charge information of the input detector signals [8].

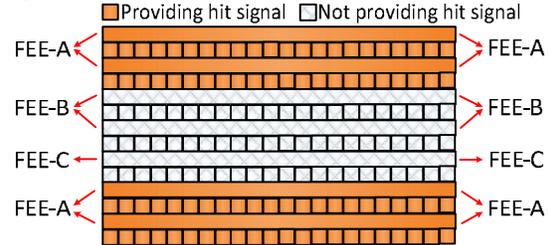

Fig. 3. Configuration for the BGO detector and FEEs

VA160 and VATA160 are the key components on FEE and they are responsible for preamplifying, shaping, sampling and holding the input analog signals and shifting out voltage signals. The detailed block diagram of VA160 is shown in Fig. 5. It contains 32 analog channels. Each channel contains an independent analog input (IN), a pre-amplifier, a semi-Gaussian pulse shaper and a sample-and-hold unit, while all 32 channels share a common calibration signal input (Cal), a differential output (outp/outn) and two 32 bits shift registers [10].

In fact, as shown in Fig. 6, VATA160 can be seen as the combination of a VA160 and a TA160, where TA160 is responsible for generating hit signals. Each channel of TA160 contains a fast shaper and a comparator, while 32-channel outputs are wired-or'ed to form a common differential hit signal output (TA/TB) [11].

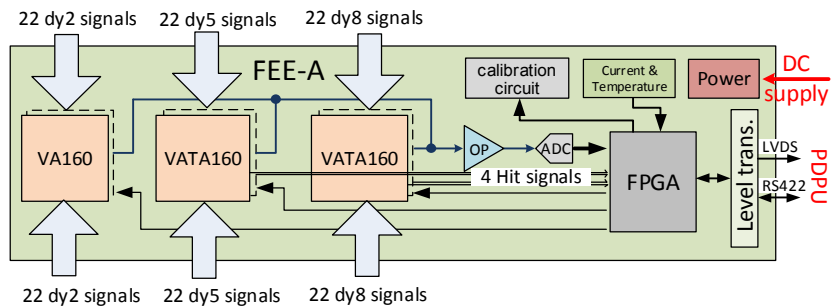

Fig. 4. The schematic of the BGO FEE

FEE contains up to 144 electronics channels and up to 6 VA chips. The calibration circuit on FEE can be used to check all charge measurement channels and hit signal generating channels remotely, by which we can learn about the status of every electronics channel and then make some adjustments in time. Futhermore, we can compensate the scientific data offline with the calibration data.

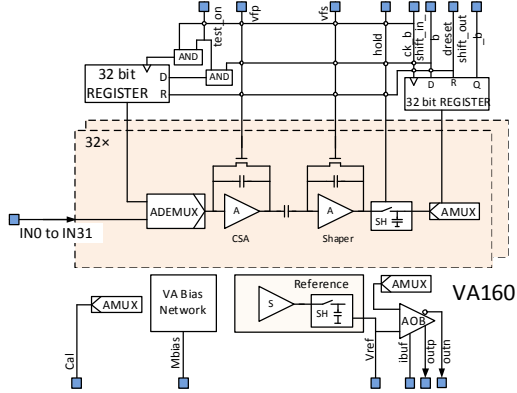

Fig. 5. Detailed block diagram of VA160

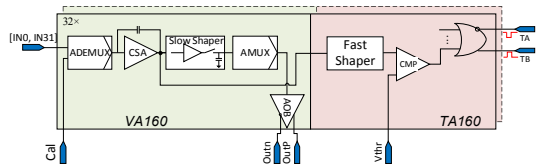

Fig. 6. Brief block diagram of VATA160

## 3 Design of the Calibration Circuit

The calibration circuit is designed with the VA160's calibration function within itself. As shown in Fig. 7, the calibration circuit on FEE mainly contains a FPGA, a DAC, an operational amplifier, an analog switch and all VA160s on FEE where FPGA controls the DAC, the analog switch and the VA160s to perform the calibration process.

### 3.1 Calibration Signal Generating

Calibration signal generating process is shown in Fig. 8. $V_{dac}$ is generated by a DAC, under control of the FPGA. As equation (1) shows, $V_{ref}$ is buffered from $V_{dac}$ through an operational amplifier. Two capacitors, C1 and C2 are used to filter out noise where C1 is 33 pF and C2 is 1nF. D1 is a diode (1N4148) used to absorb the negetive overshot of $V_{ref}$ caused by the analog switch switching from off to on [12].

In idle state, the analog switch is off and its output is pulled down to zero by R4. When the analog switch is on, a voltage equal to $V_{ref}$ is generated at the output. Controlling the analog switch switching from off to on with the FPGA, we get a rising step voltage signal, $V_{step}$. After $V_{step}$ passes through a serial capacitor, an exponential decay current pulse, $I_{cal}(t)$, is generated at the capacitor's output, which is the calibration signal. In order to obtain high-precision calibration signals, we adopt high-precision calibration capacitors (NPO capacitor, +/-1%), $C_{cal}$, and put them as close to the VA160s as possible.

$$V_{ref} = \frac{R1+R2}{R1} \times V_{dac} \quad\cdots\cdots\cdots\cdots (1)$$

$$Q_{cal} = \int_{0}^{+\infty} I_{cal}(t)\, dt \quad\cdots\cdots\cdots\cdots (2)$$

$$Q_{cal} = \int_{0}^{+\infty} \left(\frac{1}{R} V_{ref} e^{\frac{-t}{RC}}\right) dt = V_{ref} C_{cal} \quad (3)$$

In equation (1), (2) and (3), R1 and R2 is 40kΩ and 10kΩ and $C_{cal}$ is 10pF. The calibration charge input into VA160, $Q_{cal}$, is $V_{ref}*C_{cal}$, which is proportional to $V_{dac}$. $V_{dac}$ has a dynamic range of 0 ~ 2.5 volt and accordingly $V_{ref}$ has a dynamic range of 0 ~ 3.0 volt, so $Q_{cal}$ has a dynamic range of 0 ~ 30pC, which covered the VA160's input range. The other hand, the DAC has a precision of 12bits and the circuit has a low noise, which ensure the calibration circuit a precision of better than 5fC.

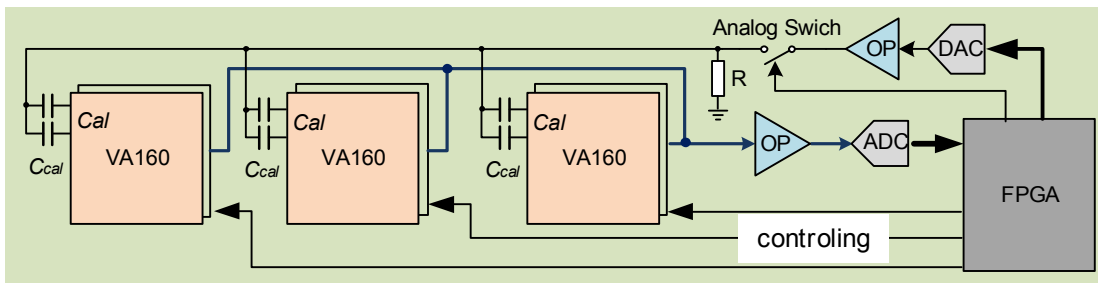

Fig. 7. Block diagram of the Calibration Circuit

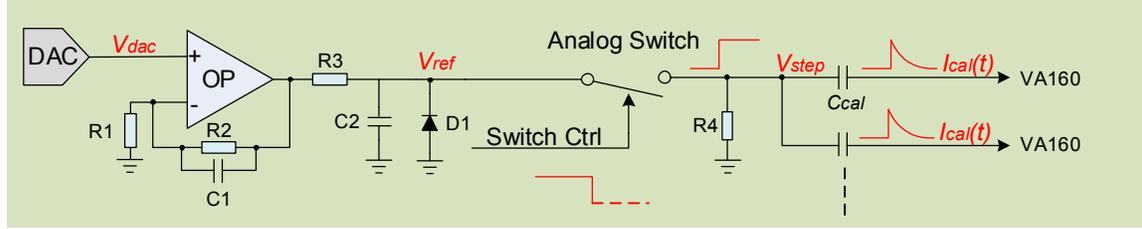

Fig. 8. Calibration signal generating circuit

### 3.2 Calibration Timing Sequence

Calibration timing sequence is shown in Fig. 9. A high level of Cali_en set the VA160s into calibration mode, then waiting for the triggers. After a fixed delay behind trigger arrives, FPGA controls the analog switch switching from off to on. And then an exponential decay signal appears on the $C_{al}$ input of the VA160s.

The calibration circuit is responsible for calibrating all channels on FEE, but it can only calibrate one channel at a time. So two 32-bit shift registers within VA160 are used to switch between channels and several VA160s are connected in a daisy chain.

All channels being calibrated, FEE will package the data and send to PDPU. After that, PDPU will send the calibration data to ground station through the satellite transmitting antenna. And then we can acquire the calibration data and analysis them offline.

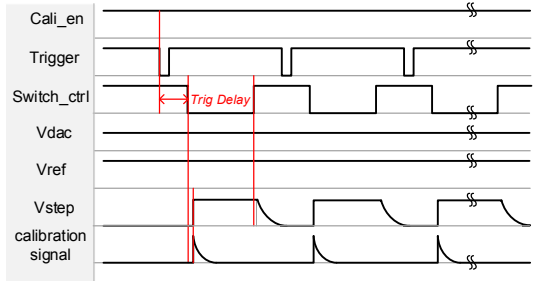

Fig. 9. Calibration timing sequence

### 3.3 Onboard Calibration Process

Onboard calibration flow chart is shown in Fig. 10. The process consists of eight steps:
(1) Power on
(2) Initialization
After powered on, all FEEs are set into normal mode of acquiring scientific data by deafault and waiting for triggers. For onboard calibration, we should set FEEs into calibration mode through command.
(3) Set FEE into calibration mode
When FEE receives a calibration command, it is set into calibration mode. FEE will set all VA160s into calibration mode and set the calibration voltage according to the command through a DAC and then wait for triggers.
(4) Trigger
When receiving a trigger, FEE starts calibration process. After a proper delay, the FPGA controls the analog switch switching from off to on to produce calibration signal. After another fixed delay (1.8 ~ 2 us), which is the signal forming time of VA160, the FPGA controls the VA160s to sample and hold the analog voltage of every channel.
(5) Analog voltages shift out
After sampled and held, the analog voltages are shifted out with differential current signal (outp, outn) in sequence from VA160s. Then the differential current signal is converted to voltage through resistance network and amplified by a instrumentation amplifier to adapt the ADC input dynamic range.
(6) Digitization
The analog voltage is digitized by a high-precision ADC (AD976) with 16-bit resolution and 200k Sps. All channels on a FEE share a common ADC and all analog voltages are digitized in sequence.
(7) Data packing
After digitized, all data of a trigger will be packed together and added a 16-bit CRC check code.
(8) Data readout
At last, the data packet will be sent out to the PDPU through the user-defined serial data bus. Only after the data packet of a trigger is sent out, FEE can receive next trigger and repeat the calibration process. After finishing all triggers at a voltage, FEE goes back to process calibration at next voltage. When calibration processes at all voltages are finished, the onboard calibration is finished.

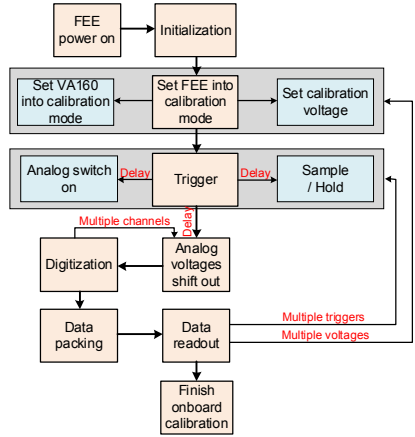

Fig. 10. Onboard calibration flow chart

## 4 Test Results

### 4.1 Linearity of the calibration signal

As shown in Fig. 11, the X-axis is the calibration DAC code and the Y-axis is voltage of the calibration signal. The voltages are measured at the analog switch output with a multimeter. It tells us that the calibration signal achieves a big dynamic range of over 3000 mV (equivalent to 30pC according to equation (3)) and a good linearity (nonlinearity is less than 0.02%). The other hand, with the DAC's high resolution of 12bit, the FEE's low electronic noise and adopting high-precision calibration capacitors with small drift (small than +/-1% from -25℃ to 55℃), we get a high resolution calibration signal better than 5fC. The calibration signal well meets the FEE's calibration requirements.

Most importantly, we test the calibration circuit in different temperature, 55℃, 25℃ and -25℃. The test results are nearly the same and their curves are coincident. The drift of the curve slope is small than 0.1% between -25℃ to 55℃. This means the calibration circuit is not sensitive to the temperature, so we can calibrate the front-end electronics accurately and real-timely.

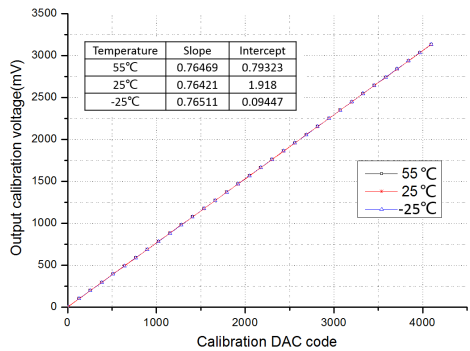

Fig. 11. Linearity of the calibration signal

### 4.2 Onboard calibration

Onboard calibration process is carried out periodically to monitor the performance of the readout channels or to compensate the parameter distortion. By calibration, we can learn about the status of every electronics channels and then make some adjustments in time [8].

The onboard calibration results of one channel are shown in Fig. 12 and Fig. 13. Fig. 12 shows the linear fitting curve with data below 12.5pC ($V_{ref}*C_{cal}$). Fig. 13 shows the RMS of all channels on a FEE at one calibration point which represents the noise of the calibration circuit. The test results tell us the calibration circuit achieves a high resolution less than 5fC.

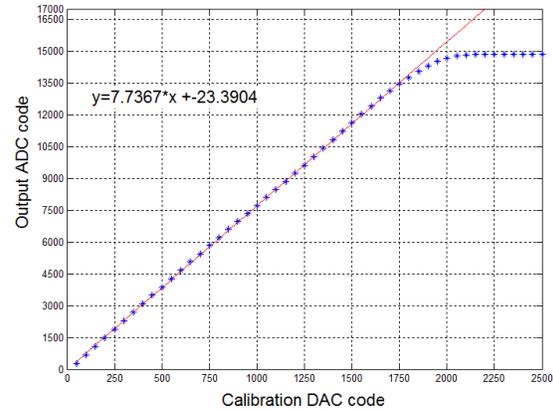

Fig. 12. Linear fitting of onboard calibration

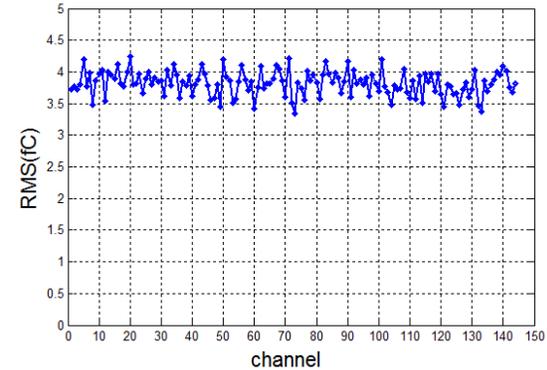

Fig. 13. RMS of onboard calibration

### 4.3 Compensating with onboard calibration in different temperature

The temperature changes rapidly along the satellite's orbiting from sunrise to sunset. For the BGO Calorimeter read-out electronics, its environmental temperature changes from -15℃ to 30℃ inside the satellite. Because the the VA160s may be interfered by temperature variation, we need to know well about the electronics' performance in different temperature. Through onboard calibrating real-timely, we can learn about FEEs' performance under all conditions and compensate the scientific data offline with the calibration results.

An external signal test with a function generator (Tek AFG3252) is carried out to simulate acquiring the charge of the dynode signals. The external voltage signal is input through a serial 10-pF capacitor to produce equivalent charge. Fig. 14 shows the test results of one channel in different temperature, -25℃, -15℃, 0℃, 25℃ and 55℃. The gain coefficient differs in different temperature where it decreases with temperature rising.

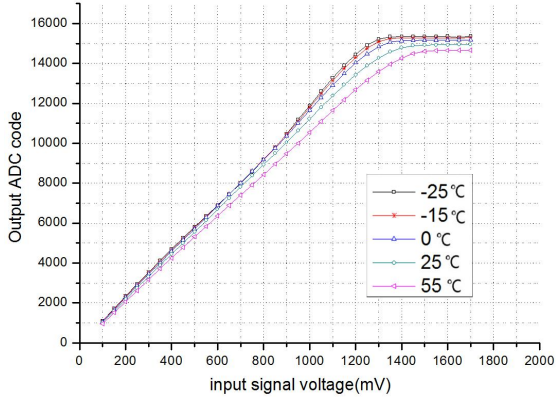

Fig. 14. External signal test in different temperature

We calibrate the FEE with onboard calibration circuit in different temperature. Calibration results of the same electronics channel in different temperature is shown in Fig. 15. The gain coefficient decreases with the temperature rising as the same trend with external signal test. So it is feasible to compensate the scientific data with onboard calibration results.

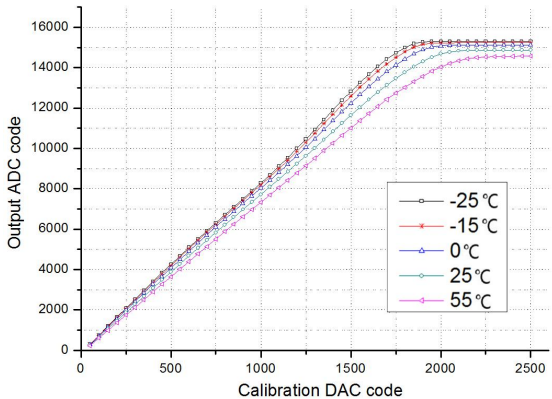

Fig. 15. Onboard calibration results in different temperature

The data are compensated according to the calibration result in room temperature (25 ℃). Firstly, we figure out the abscissas with the scientific data according to the calibration curves in different temperature, secondly we figure out the ordinates with the calculated abscissas according to calibration curves in room temperature. The calculated ordinates are compensated data which are what we want.

Fig. 16 shows the external signal test curves in different temperature after compensated with onboard calibration. The X-axis is the equivalent input signal charge ($V_{ref}*C_{cal}$) and the Y-axis is output ADC code. After compensated, all curves coincide with the one in room temperature.

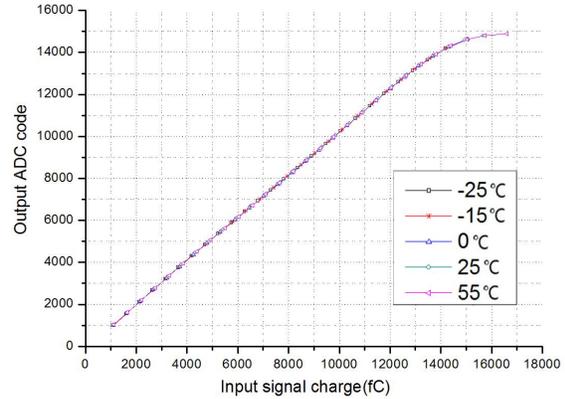

Fig. 16. External signal test results after compensated with onboard calibration

Table 1 shows the gain coefficients and INL of external signal test in different temperature before and after compensated with onboard calibration. The results show us that, after compensated, all gain coefficients are nearly equal and a better INL is achieved. Compensating the scientific data with onboard calibration results is effective.

Table 1 Gain and INL of external signal test before and after compensated

|  | Before compensated | | After compensated | |
| --- | --- | --- | --- | --- |
|  | Gain | INL | Gain | INL |
| -25℃ | 12.0 | 1.84% | 1.026 | 0.59% |
| -15℃ | 11.9 | 1.39% | 1.026 | 0.58% |
| 0℃ | 11.7 | 0.85% | 1.025 | 0.58% |
| 25℃ | 11.3 | 0.94% | 1.025 | 0.58% |
| 55℃ | 10.6 | 0.55% | 1.025 | 0.58% |

### 4.4 Onboard Calibration for Trigger Function

Another function of the calibration circuit is onboard calibration for trigger function, which is the TA160's self test to check if the hit signal generating channels work well. We use the calibration of VA160 to check the hit signal generating channels of TA160. TA160 is shown in Fig. 6. When starting TA160 calibration, we set $Q_{cal}$ (calibration signal charge) and $Q_{thr}$ (TA160's hit signal threshold) as desired and set the FEE into TA160 calibration mode by remote command. After receiving triggers, the calibration circuit sends calibration signals into VA160 and then TA160 will generate hit signals. The FPGA receives the hit signals and counters its number. When $Q_{cal}$ is higher than $Q_{thr}$, the hit signals' number will be equal to the triggers' number, else when $Q_{cal}$ is lower than $Q_{thr}$, the hit signals' number will be zero. If not so, it turns out that the corresponding hit signal generating channel goes wrong. During TA160 calibration, the FEE dosen't generate calibration data.

## 5. Conclusion

An onboard calibration circuit has been designed on the BGO FEE. It can calibrate the front-end electronics channels and help us check the status of every charge measurement channel and hit signal generating channel. We also can compensate the scientific data offline with the onboard calibration results. Test results show that the onboard calibration circuit achieves a good performance: a wide dynamic range (0~30 pC), a good resolution (5fC) and a good linearity, which well meets the requirements of FEE. Now the calibration circuit is successfully running on BGO Calorimeter read-out electronics in Qualification Model and Flight Model of DAMPE. The DAMPE satellite will be launched at the end of 2015 and the calibration circuit will play an important role in orbit.

## Acknowledgment

We would like to thank Dr. Yunlong Zhang, Dr. Zhiyong Zhang and Mr. Sicheng Wen for helpful discussions and useful suggestions. We also would like to thank all the people from DAMPE collaboration who helped make this work possible.